\documentclass[reprint,
superscriptaddress,aps,pre,twocolumn,floatfix,showpacs,amsmath,10pt]{revtex4-1}
\usepackage{amssymb,graphics}
\usepackage{setspace} 
\usepackage{graphicx}
\usepackage{hyperref}
\usepackage{xcolor}
 \usepackage{graphicx}
\usepackage{dcolumn}
\usepackage{bm}
\usepackage{xcolor}
\usepackage{graphicx}
\usepackage{graphics}
\usepackage{amsmath}
\usepackage{amssymb}
\usepackage{amsthm,amstext}
\usepackage{amsfonts}
\usepackage{mathrsfs}
\usepackage{MnSymbol}
\usepackage{mathrsfs}
\usepackage{stmaryrd}
\usepackage{latexsym}
\usepackage{bm}
\usepackage[utf8]{inputenc}
\usepackage{mdframed}
\usepackage{hyperref}
\usepackage{cancel}
\usepackage{soul}

\usepackage{yfonts}

\newcommand\by{\textbf{\emph{y}}}

\newcommand\bu{\textbf{\emph{u}}}

\newcommand\bx{\textbf{\emph{x}}}

\newcommand\bn{\textbf{\emph{n}}}

\newcommand\bk{\textbf{\emph{k}}}

\newcommand\p{\textbf{\emph{p}}}

\newcommand\B{\textbf{B}}

\renewcommand\d\delta
\newcommand\D\Delta
\newcommand\e{\varepsilon}
\newcommand\s{\sigma}
\newcommand\ph{\varphi}

\newcommand\bs{\boldsymbol{\sigma}}
\newcommand\beps{\boldsymbol{\epsilon}}

\newcommand\bet{\boldsymbol{\eta}}

\newcommand\bOmega{\boldsymbol{\Omega}}

\newcommand\curl{\text{Curl}}

\renewcommand\div{\text{div}}

\renewcommand\curl{\text{curl}}

\newcommand\Body{\mathscr{B}}

\newcommand\bbC{\mathbb{C}}

\newcommand{\trsp}{^{\hspace{-1pt}\textsf{T}\hspace{-1pt}}}

\begin{document}

\title{Printing non-Euclidean solids }

\author{Giuseppe Zurlo}%
\email{giuseppe.zurlo@nuigalway.ie}
\affiliation{School of Mathematics, Statistics and Applied Mathematics, NUI Galway, University Rd, Galway, Ireland}%

\author{Lev Truskinovsky}
\email{lev.truskinovsky@espci.fr}
\affiliation{PMMH,
CNRS - UMR 7636
PSL-ESPCI,  
10 Rue Vauquelin, 75005 Paris, France}

\date{\today}

\begin{abstract}
Geometrically  frustrated solids with non-Euclidean reference metric are ubiquitous in biology and are becoming increasingly relevant in technological applications. Often they acquire a targeted configuration of incompatibility through surface accretion of mass   as in tree growth or dam construction. We use  the mechanics of incompatible surface growth to show that geometrical frustration developing during deposition can be fine-tuned to ensure a particular behavior of the system in physiological    (or  working) conditions. As an illustration, we obtain an explicit 3D printing protocol for arteries, which guarantees  stress uniformity under inhomogeneous  loading,  and for explosive plants,  allowing a complete release of residual elastic energy with a single cut.   Interestingly, in both cases reaching the physiological target requires the  incompatibility to have a topological (global) component.

\end{abstract}

\maketitle

Externally unloaded elastic solids can be still endogenously pre-stressed by  distributed self-equilibrated force couples.  In living organisms such  pre-conditioning is a way of achieving specific targets in physiological regimes \cite{Gladman}, for instance, vegetable leafs  require residual stresses to open  \cite{Oliver} while arteries need a  pre-load to ensure transmural uniformity of hoop stress \cite{ChuFung}. Residual stresses are equally important in engineering applications, where pre-loading is used either for reinforcement or to delay the onset of  failure \cite{ResStr2,ResStrConcr}. Furthermore, challenging new applications, like the design of programmable bio-mimetic materials, depend crucially on our ability  to create complex patterns of residual stresses \cite{Ge,Kempaiah,LindNature16,Geitmann,Danescu}. 

In this Letter we  address the   question   how a particular distribution of residual stresses can be produced in a solid as a result of surface accretion of mass, a central process in both natural growth and 3D printing. 
 
The  source  of residual stresses in elastic solids is the incompatibility of the reference configuration preventing its isometric  embeding into the Euclidean space. A reference (natural) state is characterised by a metric tensor and in Euclidean solids this metric is flat \cite{Sharon2007,Efrati2009}. In ``non-Euclidean solids'', a term  apparently coined by  Poincar\'e \cite{Poincare} , the reference metric is curved, and the associated geometrical frustration manifests itself through residual stresses \cite{Hillel2016}. 

A reference curvature can be ``embedded'' into a solid by using rather well understood techniques of differential swelling, inhomogeneous thermal expansion, bulk growth and remodelling  \cite{Gladman, Santangelo2016,KupfermanPNAS,YavariThermal,C16}. Geometrical frustration can also emerge as a result of surface accretion, as in tree growth,   roll winding  and dam construction \cite{Skalak3,Archer,Correa,Gibson,Arut}.  In the case of surface growth, the relation between the physics of deposition and the resulting incompatibility is implicit and it is not clear which accretion protocol  leads to  a desired distribution of residual stresses. 

Surface growth is often modeled in a holonomic format of  elastically  coherent phase transitions, that take place without  generation of incompatibility \cite{Ganghoffer2012,Ciarletta2013}. A more general, non-holonomic approach  should allow for the  incompatibility to be  acquired  at the moment of the creation of a new continuum particles, see for instance \cite{BG,Trincher,Gamba}.  In this Letter we consider a general inverse problem of this type and view  the deposition stress as a tensorial control parameter. We  find an explicit link between the implemented deposition strategy and the resulting incompatibility. The obtained relations not only reveal the mechanisms of biological adaptation associated with surface growth but  can  also guide additive manufacturing of programmable meta-materials. 
 
To illustrate the general theory, we study in detail the process of artificial 3D printing of arteries.  Presently, the  circumferential wrapping of sheets of living cells  is used to reproduce their natural layered structure \cite{NatureMain2015,Lheureux2009,Lheureux2012}. However, in physiological conditions  the resulting transmural stress distribution is far from realistic  \cite{Zaucha,VarnerNelson2014}. Instead, our approach allows  reaching the target physiological state precisely, and we show that the proposed strategy is compatible with available manufacturing technologies.  As another illustration,  we design a growth protocol  ensuring that a single cut in a  hollow cylinder results in a complete release of the residually stored elastic energy. This prototypical problem is relevant  for the understanding of explosive seed dispersal and other functions of plant actuators \cite{Hofhuis2016,Shahaf2011,Xuxu2016}. Quite remarkably, we find that in  both biological examples  the crucial role is played by  a  global contribution to incompatibility usually associated  with topological defects in crystals (disclinations) \cite{Anthony}.

Consider a body $\Body_p$ that in physiological (or  working)  conditions is subjected to tractions ${\bf s}_p$ applied on its boundary $\omega_p$. The symmetric physiological stress field  $\bs_p(\bx)$ must satisfy the equilibrium conditions 
\begin{equation}
\label{1}
\div\,\bs_p={\bf 0} \quad\text{in}\quad\Body_p, \quad\quad\quad
 \bs_p \bn =  {\bf s}_p\quad\text{on}\quad \omega_p
\end{equation} 
where $\bn$ is the outward normal. In the absence of additional equations, the stress remains under-determined. If the deformation is Euclidean  (compatible, neighbor preserving, elastic)  the problem can be closed by supplementing \eqref{1} with relations expressing the stress in terms of the gradients of vector-valued displacements.   This leads to a distribution of  stresses that can be altered only by changing the shape of the body   or by varying the elastic properties of the material. An alternative  way  to control the stress state of the body and its deformed shape,  apparently favored by biological systems,  is to give up the   compatibility and employ inelastic deformations.   

Suppose (for simplicity) that the Hooke's law still holds for incremental deformations and define the (linear) elastic strain $\beps_p=\bbC^{-1}\bs_p$, where $\bbC$ is the elasticity tensor. The signature of the non-Euclidean character of the stress is the nonzero incompatibility $\bet_p=\curl\,\curl\beps_p$,  the linear counterpart of the reference Riemann curvature, which satisfies the   Bianchi identities $\div\,\bet_p={\bf 0}$. If a target incompatibility $\bet_p$ is prescribed, its three independent components remain unconstrained by \eqref{1} and can be used to ``engineer'' a particular physiological state of stress.    Embedding a  strain incompatibility into the solid can be viewed as a way of ``programming''  the material, whose ultimate performances will   depend on the loads and the shape of the body, e.g.  \cite{Danescu}.
 
Note that $\bet_p$ should be understood in the sense of distributions \cite{VGvolterra} because the target incompatibility may   contain both diffuse and singular contributions. Furthermore, singular defect lines may have a global effect if  they carry topological charges  characterized by  the nonzero Burgers   $\B_p(\bx_o)=\lim_{h\rightarrow 0} \int_{D_h}\by\times\bet_p\trsp\bn\,da$ and the Frank  $\bOmega_p(\bx_o)= \lim_{h\rightarrow 0}\int_{D_h}\bet_p\trsp\bn\,da$ vectors \cite{Volterra,Weingarten, KupfermanPNAS,SM} where  $\by$ is the position vector of points of an asymptotically  shrinking oriented disk $D_h$ of diameter $h$, enclosing the singular point $\bx_o$ \cite{note}. Since the associated residual stresses  cannot be removed by cutting  singular lines out of the body,  in non-simply connected bodies such topological charges may be located outside the domain $\Body_p$. 

To model a non-holonomically growing body we introduce a sequence of (incremental) configurations $\Body (t)$.   The time-like parameter $t$ changes in the interval  $(t_i, t_f)$, denoting the  beginning and the end of the accretion process. In particular, $\Body (t_f)=\Body_p$. We denote by $\tau(\bx)$ the instant when the accreting surface $\omega(t)$, whose evolution is assumed to be known, passes through a point $\bx$. 

\begin{figure}[!th]
\includegraphics[scale=.16]{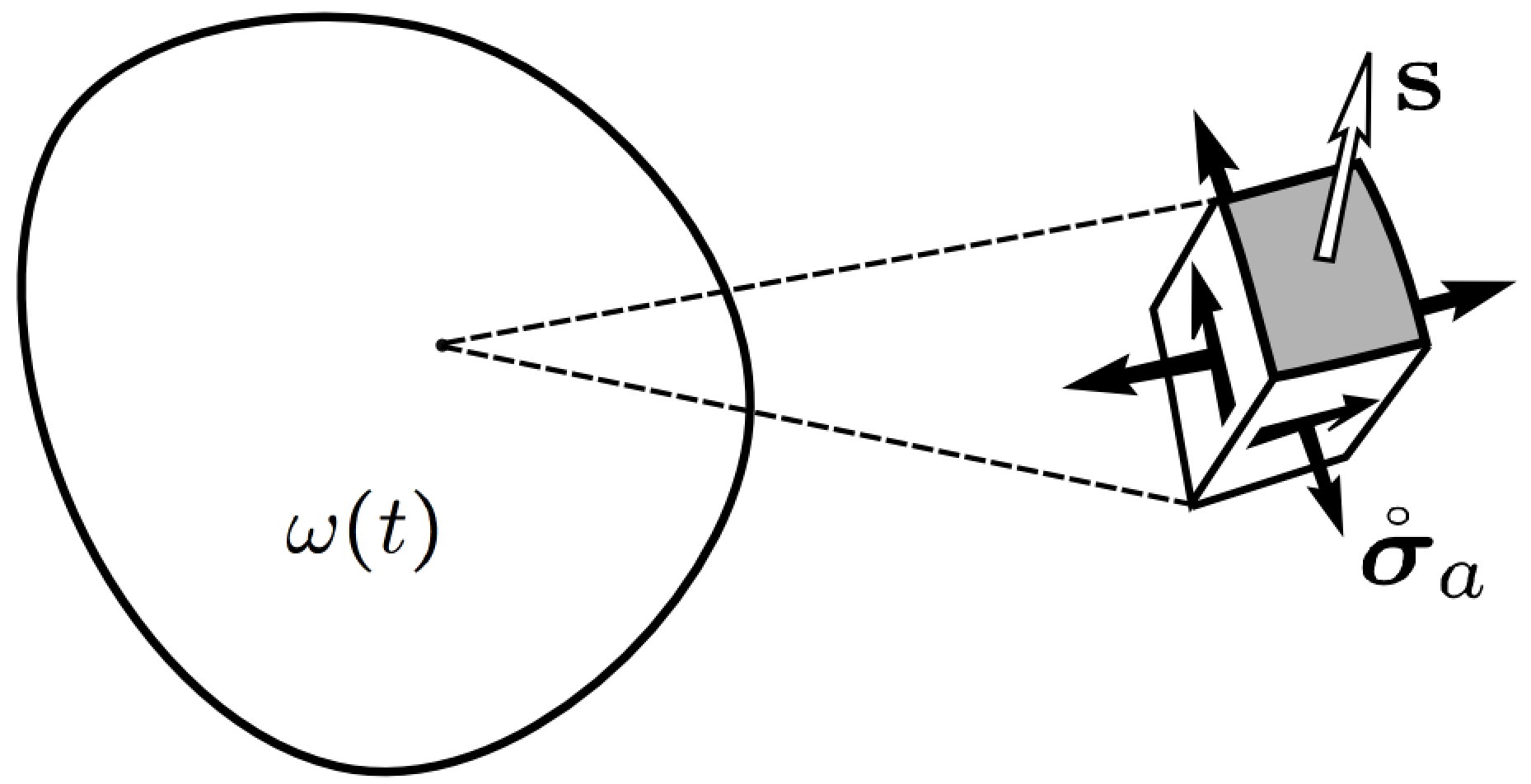}
\caption{\label{Scheme}\footnotesize{A sketch of the deposition  surface $\omega(t)$ showing an infinitesimal element subjected to   external tractions ${\bf s}$ (three components) and controlled by active surface stresses $\mathring{\bs}_a$ (three components).}}
\end{figure}

At $t\geq\tau(\bx)$ the stress tensor in the growing body is
$\bs(\bx,t) = \mathring{\bs}(\bx) + \int_{\tau(\bx)}^{t}\dot\bs(\bx,s)\,ds$, 
where  the deposition  stress $\mathring{\bs}$ can be further decomposed into a sum of two (rank-two) contributions: $\mathring{\bs}(\bx)=\p(\bx)+\mathring{\bs}_a(\bx)$,  see Fig.\ref{Scheme}.  While the ``passive'' contribution $ \p ={\bf s}\otimes\bn + \bn\otimes{\bf s}-({\bf s}\cdot\bn)\bn\otimes\bn$ is fully defined by the 
 tractions ${\bf s} = \bs\bn$, the ``active''  surface stress $\mathring{\bs}_a $, satisfying $\mathring{\bs}_a\bn=\mathring{\bs}_a\trsp\bn=\textbf{0}$, carries three independent degrees of freedom that  can be used to ``implant defects'' in the upcoming layers.  

The breakdown of  $\mathring{\bs}$ into active and passive contributions is somewhat arbitrary, because in both natural and technological conditions the three  components of $ \p$ may  also   play the  role of active agents. In some other cases the implied freedom simply disappears: during  coherent structural  transformations the whole stress $\mathring{\bs}$ is determined by the evolution laws of the  interface \cite{Trusk83,Ciarletta2013}; for solidification, it is natural to assume that the active term adjusts to ensure that $\mathring{\bs} $ is hydrostatic \cite{KF}; for surface growth in plants and in some additive manufacturing processes, the adhering layer can be treated as a pre-stressed elastic membrane, whose state of deformation is controlled by the requirement of equilibrium under suitable anchoring and adhesion conditions \cite{Manz3,GorielyKuhl}. Here we neglect the potential  constraints imposed by the  deposition mechanism and we treat the independent components of $\mathring{\bs}_a$ as free control parameters. 

Since  equilibrium  must hold at each stage of growth, we must have  $\div\,\bs=\textbf{0}$ in $\Body(t)$ and $\bs =\mathring{\bs}$ on $\omega(t)$; note that the ``whole''  stress tensor is prescribed on the growing surface. We   assume that inelastic phenomena leading to the accumulation of incompatibilities take place only at the instant of the deposition. Away from the accreting surface the incremental behavior is assumed to be linearly elastic. If we neglect the effect of prestress on the incremental behavior \cite{Ho}, we can write  $\dot\bs=\bbC\dot\beps$, where  $\dot\beps=(\nabla\dot\bu+\nabla\dot\bu\trsp)/2$ and  $\dot\bu$ is  the incremental displacement. At a given $t$ we can formulate an   incremental problem of elasticity theory in the form \cite{Trincher,SM}
\begin{equation}
\label{Trincher1}
\left\{
\begin{array}{llll}
\div\,\dot\bs=\textbf{0} & \quad & \text{in} & \Body(t)\\
\dot\bs \bn = |\nabla\tau|^{-1}\div(\mathring\bs_a+ \p) & \quad & \text{on} &  \omega(t).
\end{array}
\right.
\end{equation} 
By  solving a  sequence of such problems we obtain the  total elastic strain $\beps(\bx,t) = \mathring\beps(\bx)+\int_{\tau(x)}^t\dot\beps(\bx,s)\,ds$, where the accreting contribution $\mathring\beps=\bbC^{-1}(\mathring{\bs}_a+\p)$, which is also assumed to be small, is generally incompatible. If we now require  that   
  the final incompatibility  $\bet(\bx,t_f) = \curl\,\curl\beps(\bx,t_f)$ equals its target physiological value $\bet_p(\bx)$, we obtain a constraint on the instantaneous incompatibility of the arriving material $\mathring\bet  = \curl\,\curl\mathring\beps $ in the form 
\begin{equation}\label{general1}
\mathring\bet-\nabla\tau\times\left[\curl\,\dot\beps\right]_{\omega}\trsp - \curl\left[\nabla\tau\times\dot\beps_{\omega}\right]=\bet_p\quad\text{in}\quad\Body(t_f). 
\end{equation}
Here we used the notation  $A_{\omega}(\bx):=A(\bx,\tau(\bx))$, see \cite{SM} for details. Since in \eqref{general1} , $\dot\beps(\bx,t)$   implicitly depends  on $\mathring{\bs}_a(\bx)$ through \eqref{Trincher1},  we now have a nonlocal  relation between the three independent controls of $\mathring\bs_a $ and the  three independent targets in  $\bet_p$. In the cases when these relations should be understood in the sense of distributions, we implicitly require  that in each singular point $\bx_0$, $\B(\bx_0,t_f) = \B_p(\bx_0)$ and $\bOmega(\bx_0, t_f)=\bOmega_p(\bx_0)$. Equation \eqref{general1}, which is the main result of this  Letter,  defines the deposition strategy that ensures the attainment of a desired stress distribution in physiological conditions.   

\begin{figure}[!th]
\includegraphics[scale=.18]{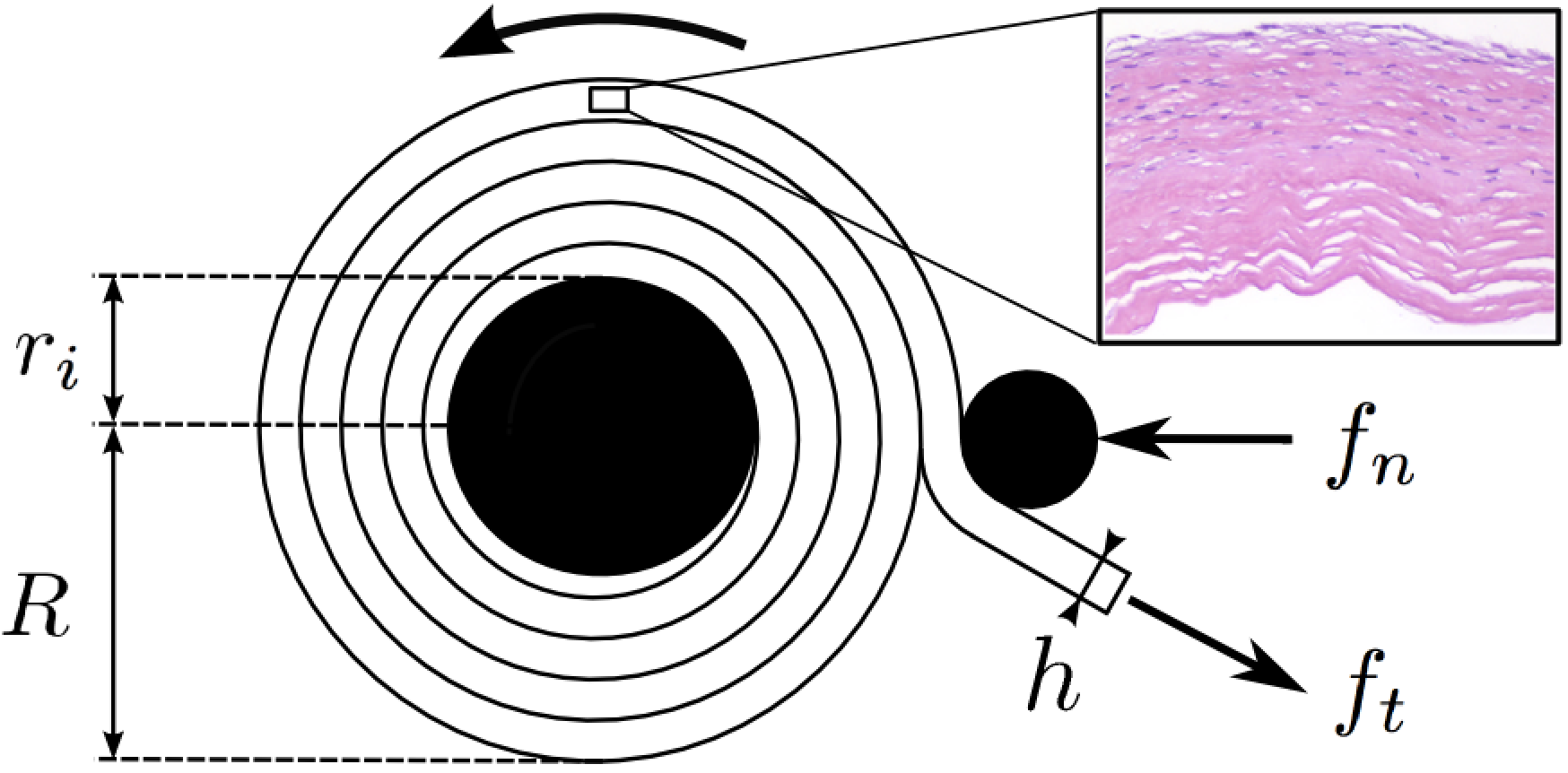}
\caption{\label{Winding}A sketch of the winding process during artificial manufacturing of an artery, showing the deposition of a layer of thickness $h$ subjected to a ``passive'' force $  f_n$ and an ``active'' force $  f_t$.  Our ``microscopic'' formulation corresponds to the limit $h \to 0$,  $  f_t \to 0$, while $  f_t / h$ remains finite.  Inset: the cross-section of an artery grown by winding layers of mesenchymal cells (courtesy of  \cite{Lheureux2009}).}
\end{figure}

As an  illustration,  consider  the process of layered manufacturing of an artery \cite{NatureMain2015,Lheureux2012}. For simplicity, the artery will   be modeled as a hollow, infinitely long cylinder loaded in plane strain. This is equivalent to replacing a cylinder by a disk which makes the problem two-dimensional and fully explicit. We assume that the deposition  starts on a rigid mandrel of radius $r_i$ and that the  disk  grows outwards till the final (physiological)  radius $r_f$ is reached. It is convenient to use as a  time-like parameter $R=R(t)$, representing the current radius of the accreting surface (line),  so that $\tau(R)\equiv R$. Intermediate configurations of the artery are then represented by $r_i\leq r\leq R\leq r_f$. See Fig.\ref{Winding}  for  the ``macroscopic'' rendering of this process: our ``microscopic''  formulation corresponds to the limit when the thickness of the attached layers $h \to 0$.

If the  elastic solid is isotropic and  the deposition strategy respects polar symmetry, the incremental displacement reduces to its radial component $\dot u(r,R)$. In this case  the incremental radial and hoop strains are $\dot\e_r = \partial_r\dot u$ and $\dot\e_{\theta}=\dot u/r$, respectively, where the superposed dot denotes $\partial/\partial R$. The  incremental stress rates   $\dot\s_{r/\theta}=2\mu\dot\e_{r/\theta}+\lambda(\dot\e_r+\dot\e_{\theta})$, where $\lambda$ and $\mu$ are the Lam\'e moduli, must satisfy the equilibrium equation  $\partial_r\dot\s_r + \left(\dot\s_r - \dot\s_{\theta}\right)/r=0$. 

Observe that  the applied tractions  have only radial component $s(R)$, and that the surface component of the deposition stress is  fully characterized by its  hoop component $\mathring\s_{a_{\theta}}(R) \sim f_t/h$, see  Fig.\ref{Winding}. Then (\ref{Trincher1})$_2$    reduces to $\dot\s_r (R,R)= g(R)$ where $g(R) = s'(R) + \left(s(R) - \mathring\s_{a_{\theta}}(R)\right)/R$. Assuming that displacements are fixed on the rigid   mandrel, $\dot u(r_i,R) = 0$, we obtain an explicit solution of the incremental problem
\begin{equation}\label{displ}
\dot u(r,R) = \frac{R^2 f(R)}{\mu\,r_i^2 + (\mu+\lambda) R^2}\frac{r^2-r_i^2}{2 r}. 
\end{equation}
 
\noindent To specify the  deposition protocol $\mathring\s_{a_{\theta}}(r)$ we need to satisfy  \eqref{general1} and match the target  topological constraints imposed through  $\B_p$ and $\bOmega_p$. In view of our symmetry assumptions, the strain incompatibility tensor reduces to $\bet=\eta(r)\bk\otimes\bk$, where the unit vector $\bk$ is aligned with the cylinder axis, $\eta={\e_{\theta}}''+(2{\e_{\theta}}'-{\e_r}')/r$ and  $'=\partial/\partial r$. Eq. \eqref{general1}  reduces to $\mathring\eta(R) - (R\,\dot\e_{\theta}(R,R))'/R = \eta_p(R)$ and    conditions on a potential line singularity at $r=0$ take the form $\B_p=\textbf{0}$ and $\bOmega_p=2\pi\ph(r_i)r_i\bk$. Here we introduce the function $\ph(r)={\e_{\theta}}'+({\e_{\theta}}-\e_r)/r$, see \cite{SM} for additional details. Since $\eta={(\ph\,r)}'/r$, we can  recast \eqref{general1} in the form  
$
\mathring\ph(R) - \dot\e_{\theta}(R,R) = \ph_p(R), 
$
where $\mathring\ph(r)$ refers to the arriving material and $\ph_p(r)$ to the physiological target state, while  $\dot\e_{\theta}(r,R)$ is calculated from \eqref{displ}.    Note that if $\ph_p(r_i)\neq0$, the target incompatibility has a nonzero  topological (global) component.

If we now assume for determinacy that $s(R)=0$ and   $\mathring\s_{a_{\theta}}(r_i)=0$,  we can express  the function $ \mathring\ph(R)$ in terms of $\ph_p(r)$.  This gives the  desired deposition strategy  securing the attainment of a generic incompatibility: 
\begin{equation}\label{sthetafinal}
\mathring\s_{a_{\theta}}(R) = \frac{4\mu(\mu+\lambda)\int_{r_i}^{R}\left(\mu r_i^2 + (\mu+\lambda) r^2\right)\ph_p(r)\,dr}{(2\mu+\lambda)(\mu r_i^2 + (\mu+\lambda)R^2)}. 
\end{equation}

\begin{figure}[!th]
\includegraphics[scale=.23]{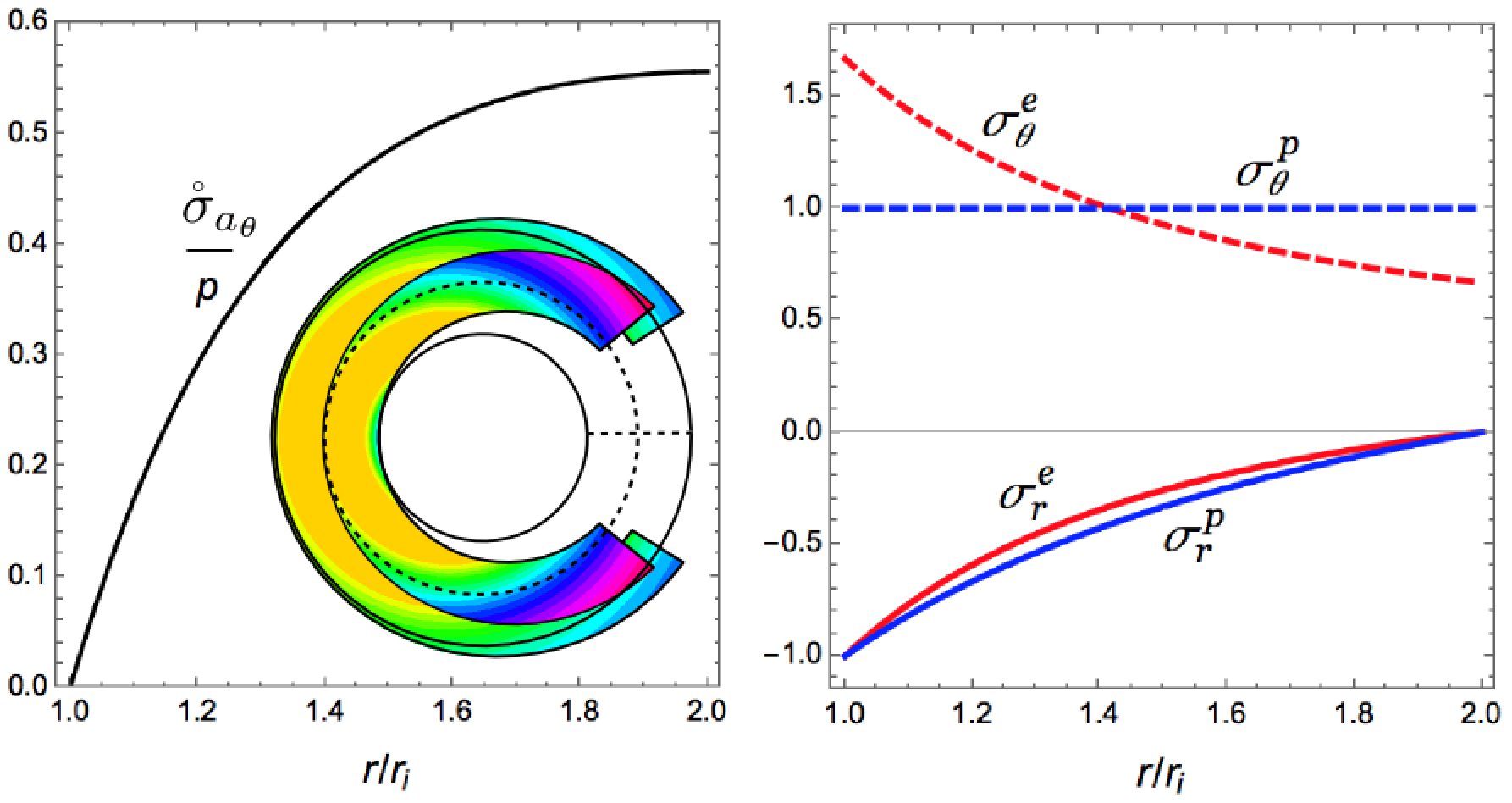}
\caption{\label{strategy1}\footnotesize{{\it Left:} Deposition strategy guaranteeing transmural uniformity of hoop stress in physiological conditions. The inset shows a FEM simulation illustrating the  stress norm (yellow=0, magenta=max) and the displacement field  resulting from cutting the disk along the dashed lines \cite{SM}. {\it Right:} Purely elastic $(\s_r^e,\s_{\theta}^e)$ vs growth induced (inelastic) $(\s_r^p,\s_{\theta}^p)$ stress distributions, for an internal pressure $p=1$.}}
\end{figure}

 For arteries, the physiological state is characterised by a finite internal pressure $p$ acting on $r=r_i$ and a much smaller external pressure acting on $r=r_f$, which we assume to be equal to zero.  Under these conditions,  stresses in a purely elastic tube would be transmurally inhomogeneous (Fig.\ref{strategy1}), which is incompatible with experiments \cite{Holzapfel2007} pointing towards homogeneity of  the hoop stress  \cite{ChuFung}. To find the   physiologically justified incompatibility which guarantees that   ${\s^p_{\theta}}'=0$,   we combine this target condition with the equilibrium equation ${\s_r^p}'+(\s_r^p-\s_{\theta}^p)/r=0$ and   obtain that $\s^p_r = -p(r_f - r)r_i/(r(r_f - r_i))$ and $\s_{\theta}^p=p\,r_i/(r_f-r_i)$. We  can now compute the function $\ph_p(r)$ and substitute it into \eqref{sthetafinal}. The resulting  deposition strategy 
\begin{equation}
\frac{\mathring\s_{a_{\theta}}(R)}{p} = \frac{r_i\,r_f (R-r_i)(\mu r_i + (\mu+\lambda)R)}{(r_f-r_i)R (\mu r_i^2 + (\mu+\lambda)R^2)} 
\end{equation}

\noindent is illustrated  in Fig.\ref{strategy1}. 

Note that the singular component of the incompatibility does not vanish  since $\ph_p(r_i)=p(2\mu+\lambda)r_f/\left(4\mu(\mu+\lambda)(r_f-r_i)r_i\right)$; the physiological conditions then require the presence of  a  ``ghost'' wedge disclination (or its diffuse analog) aligned with the axis of the artery.  Since the non-singular part of the incompatibility is also different from zero, the  residual stresses cannot be relaxed by a single longitudinal cut turning the cylinder into a simply connected domain. This is consistent with experiments on arteries \cite{Holzapfel2007}, showing that the internal layer ({\it media}) has a greater opening angle than the external layer ({\it adventitia}), see figures in \cite{SM}. Such behavior is also reproduced by our FEM simulations (Fig.\ref{strategy1}) for a disk manufactured following the proposed strategy, see \cite{SM} for  details on numerics. 
\begin{figure}[!th]
\includegraphics[scale=.23]{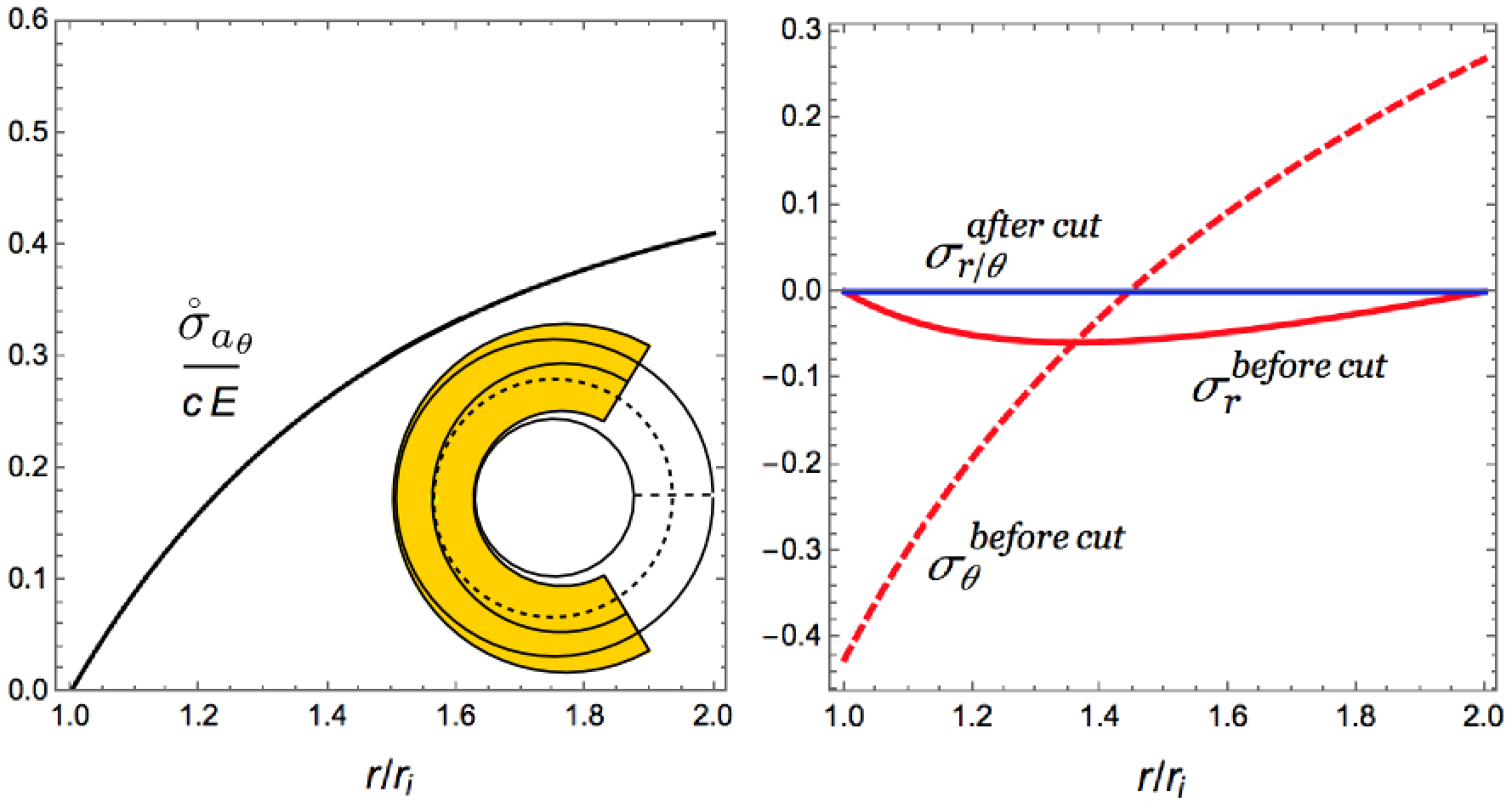}
\caption{\label{strategy2}\footnotesize{{\it Left:} Deposition strategy guaranteeing that the stored elastic energy is completely released by a single cut. The inset shows a FEM simulation of the displacements field resulting from cutting the disk along the  dashed lines. The stresses are everywhere zero after the first cut \cite{SM}. {\it Right:} Stresses in the disk before and after radial cut;  $cE=1$.}}
\end{figure}

As a second illustration, consider a rather  different physiological target  that may be relevant for explosive plants \cite{Hofhuis2016}.  Keeping the same geometry as in the case of arteries, we demand that the distribution of incompatibility is such  that the stored elastic energy due to residual stresses is fully released with a single global cut. This requirement will be met if we grow a hollow tube with $\eta_p=0$ in the bulk and nonzero $\Omega_p$. To this end  we must choose $\ph_p(r)=c/r$, with $c$ a constant characterising the magnitude of the stored/released energy.  The resulting singular incompatibility field  can be interpreted as a Volterra's wedge disclination with an opening angle $\Omega_p=2\pi c$ \cite{SM,VGvolterra,Anthony}. If we now substitute  this incompatibility into \eqref{sthetafinal}, we obtain that the  deposition strategy
\begin{equation}\label{Explosive}
\frac{\mathring\s_{a_{\theta}}(R)}{c} = -\frac{2\mu(\lambda+\mu)}{\lambda+2\mu}\frac{(\mu+\lambda)(r_i^2-R^2)+2r_i^2\mu\log(r_i/R)}{\mu r_i^2 + (\mu+\lambda)R^2}.
\end{equation}
It is illustrated in Fig.\ref{strategy2}, where we also show by FEM  simulation that a single longitudinal slicing  of  a pre-stressed cylinder with this (purely singular) incompatibility indeed leads to a complete release of the residual stresses, and that subsequent orthogonal slicing does not produce additional relaxation.

To illustrate yet another type of  protocols where both tensorial components of the   deposition stress, $\p(\bx)$ and $\mathring{\bs}_a(\bx)$, play an active role, we  assume that the newly arriving continuum particles are hydrostatically pre-stressed, with the control parameter $\pi$ representing negative pressure. In the same geometrical setting as above we get  $\pi(R)=s(R) = \mathring\s_{a_{\theta}}(R)$ and,  following an almost identical  line of reasoning, we obtain that  under such boundary/deposition  conditions the strain distribution characterized by a generic  function  $\ph_p(R)$ can be reached  if we use the  protocol
\begin{equation}\label{Pifinal}
\pi(R) = \frac{2(\mu+\lambda)}{r_i^2(2\mu+\lambda)}\int_{r_i}^{R}\left(\mu r_i^2 + (\mu+\lambda) r^2\right)\ph_p(r)\,dr. 
\end{equation}
Clearly both targets considered above,  hoop stress uniformity and a complete release of energy with a single cut, can be achieved in this framework as well.  An important example of such   hydrostatic ``printing'' is the crystallisation in a closed container, where the inhomogeneity of the deposition pressure  is ensured by the finite compressibility of the melt, e.g. \cite{KF}.

In conclusion, we outlined a new theoretical  framework for controlled  incompatible surface growth and obtained   explicit relations that can be used to guide additive manufacturing.  Acquiring  an ability to generate complex  patterns of residual stresses is a  crucial step in both biological evolution and  the design  of bio-mimetic  meta-materials. The proposed surface deposition strategy promises to bring a combination of unprecedented level of control, together with the ability to handle arbitrarily complex geometries. Future studies are needed to tailor our general theory to specific deposition technologies \cite{Kong}, to extend it to finite strains \cite{SozioYavari}, and to develop  an energetic framework coupling  the velocity of the accretion front with the corresponding driving forces \cite{DiCarloSurfBulk}.
 
\emph{Acknowledgments.} The authors thank M.Destrade, P.Recho and B.Shoikhet for helpful discussions.  G.Z. was supported by  the ERC Marie Curie Fellowship, INdAM $\&$ GNFM;  L.T. was supported by the PSL grant  CRITBIO.


\begin{thebibliography}{99}
\bibitem{note} Since  $\B_p$  depends on the choice of the coordinate system, the three degrees of freedom associated with $\bOmega_p\neq0$ can   serve as a complete invariant characterization of the global effect  of a defect  line.
\end{thebibliography}
\end{document}